\documentclass[times,twocolumn,final]{elsarticle}
\usepackage{medima}
\usepackage{framed,multirow}

\usepackage{amssymb}
\usepackage{latexsym}

\usepackage{url}
\usepackage{xcolor}

\usepackage{hyperref}

\usepackage{amsmath}
\usepackage{adjustbox}

\definecolor{newcolor}{rgb}{.8,.349,.1}

\journal{Medical Image Analysis}

\begin{document}

\verso{Bokhorst \textit{et~al.}}

\begin{frontmatter}

\title{Automated risk classification of colon biopsies based on semantic segmentation of histopathology images}

\author[1]{John-Melle \snm{Bokhorst}\corref{cor1}}
\cortext[cor1]{Corresponding author: 
  Tel.: +31-243614314 ;  
  E-mail: John-Melle.Bokhorst@radboudumc.nl;}
\author[1]{Iris D. \snm{Nagtegaal}}
\author[2]{Filippo \snm{Fraggetta}}
\author[2]{Simona \snm{Vatrano}}
\author[3]{Wilma \snm{Mesker}}
\author[4]{Michael \snm{Vieth}}
\author[1,5]{Jeroen \snm{van der Laak}}
\author[1]{Francesco \snm{Ciompi}}

\address[1]{Radboud University Medical Center Department of pathology, Nijmegen, The Netherlands}
\address[2]{Dipartimento di Anatomia Patologica, Ospedale Cannizzaro, Catania, Italy} 
\address[3]{Leids Universitair Medisch Centrum, Leiden, The Netherlands} 
\address[4]{University of Bayreuth, Bayreuth, Germany} 
\address[5]{Center for Medical Image Science and Visualization, Linköping University, Linköping, Sweden}

% \author[1]{Given-name1 \snm{Surname1}\corref{cor1}}
% \cortext[cor1]{Corresponding author: 
%   Tel.: +0-000-000-0000;  
%   fax: +0-000-000-0000;}
% \author[1]{Given-name2 \snm{Surname2}\fnref{fn1}}
% \fntext[fn1]{This is author footnote for second author.}
% \author[2]{Given-name3 \snm{Surname3}}
% %% Third author's email
% \ead{author3@author.com}
% \author[2]{Given-name4 \snm{Surname4}}

% \address[1]{Affiliation 1, Address, City and Postal Code, Country}
% \address[2]{Affiliation 2, Address, City and Postal Code, Country}

% \received{1 May 2013}
% \finalform{10 May 2013}
% \accepted{13 May 2013}
% \availableonline{15 May 2013}
% \communicated{S. Sarkar}

\begin{abstract}
%%%
Artificial Intelligence (AI) can potentially support histopathologists in the diagnosis of a broad spectrum of cancer types. 
In colorectal cancer (CRC), AI can alleviate the laborious task of characterization and reporting on resected biopsies, including polyps, the numbers of which are increasing as a result of  CRC population screening programs, ongoing in many countries all around the globe. 
Here, we present an approach to address two major challenges in automated assessment of CRC histopathology whole-slide images. 
First, we present an AI-based method to segment multiple tissue compartments in the H\&E-stained whole-slide image, which provides a different, more perceptible picture of tissue morphology and composition. 
We test and compare a panel of state-of-the-art loss functions available for segmentation models, and provide indications about their use in histopathology image segmentation, based on the analysis of a) a multi-centric cohort of CRC cases from five medical centers in the Netherlands and Germany, and b) two publicly available datasets on segmentation in CRC. 
Second, we use the best performing AI model as the basis for a computer-aided diagnosis system (CAD) that classifies colon biopsies into four main categories that are relevant pathologically.
We report the performance of this system on an independent cohort of more than 1,000 patients.
The results show the potential of such an AI-based system to assist pathologists in diagnosis of CRC in the context of population screening. 
We have made the segmentation model available for research use on https://grand-challenge.org/algorithms/colon-tissue-segmentation/. 

%%%%
\end{abstract}

\begin{keyword}
% %% or \MSC[2008] code \sep code (2000 is the default)
% \MSC 41A05\sep 41A10\sep 65D05\sep 65D17
%% Keywords
\KWD Computational Pathology\sep Colon Cancer\sep Colon Biopsies \sep Colon Polyps \sep Deep learning \sep Segmentation \sep Classification 
\end{keyword}

\end{frontmatter}

\section{Introduction}

% introduction on CRC screening and needs for AI
Colorectal cancer (CRC) is the third most commonly occurring cancer in men and the second in women, and is expected to affect more than 2.2 million new cases and cause 1.1 million deaths by 2030 \citep{eusc17}.
Population \emph{screening} is the key to the early detection of CRC and greatly increases the chances for successful treatment, reducing morbidity and mortality. 
In Europe, more than 110 million people are being targeted by annual screening programs, where 5\% of the participants require follow-up colonoscopy assessment, often resulting in taking biopsies and possibly polypectomy.
As a result, histopathological centres in Europe are facing a massive increase in tissue for processing and diagnosis. 

% classification
\subsection{Biopsy classification}

During diagnosis, pathologists analyze the tissue morphology and use it to classify a case by recognizing patterns of cell characteristics of 1) \emph{early invasive cancers}, i.e., malignant biopsies that consist of tumor cells invading through the muscularis mucosae into the underlying submucosa (pT1 tumor), 2) \emph{high-risk precancerous lesions}, with adenomatous (i.e., clonal lesions that show at least low-grade dysplasia), and serrated polyps (recently recognized as important as the adenoma-carcinoma pathway), including \emph{hyperplastic biopsy}, as the main representatives. 
In this context, automation of tissue recognition and classification of (pre-)malignant biopsies can potentially relieve the daily clinical routine workload of pathologists, e.g., by pre-screening cases.
The advent of digital pathology allows the development of computer algorithms (often referred to as computational pathology) to aid pathologists in classifying biopsies.
To the best of our knowledge, only one previous study \citep{korb17} directly addressed the problem of automatic classification of colon biopsies whole-slide images, but it is focused on five non-neoplastic conditions, namely hyperplastic, sessile serrated, traditional serrated, tubular, and tubulovillous/villous, which are included in the US Multisociety Task Force guidelines for CRC risk assessment and surveillance.
Other authors \citep{xu20b} have focused on the classification of regions of interest (i.e., patches) of WSI as containing tumor or non-tumor tissue without attempting to assess entire colon biopsies.

% figure: grades
\begin{figure}
  \centering
  \includegraphics[width=1.0\linewidth]{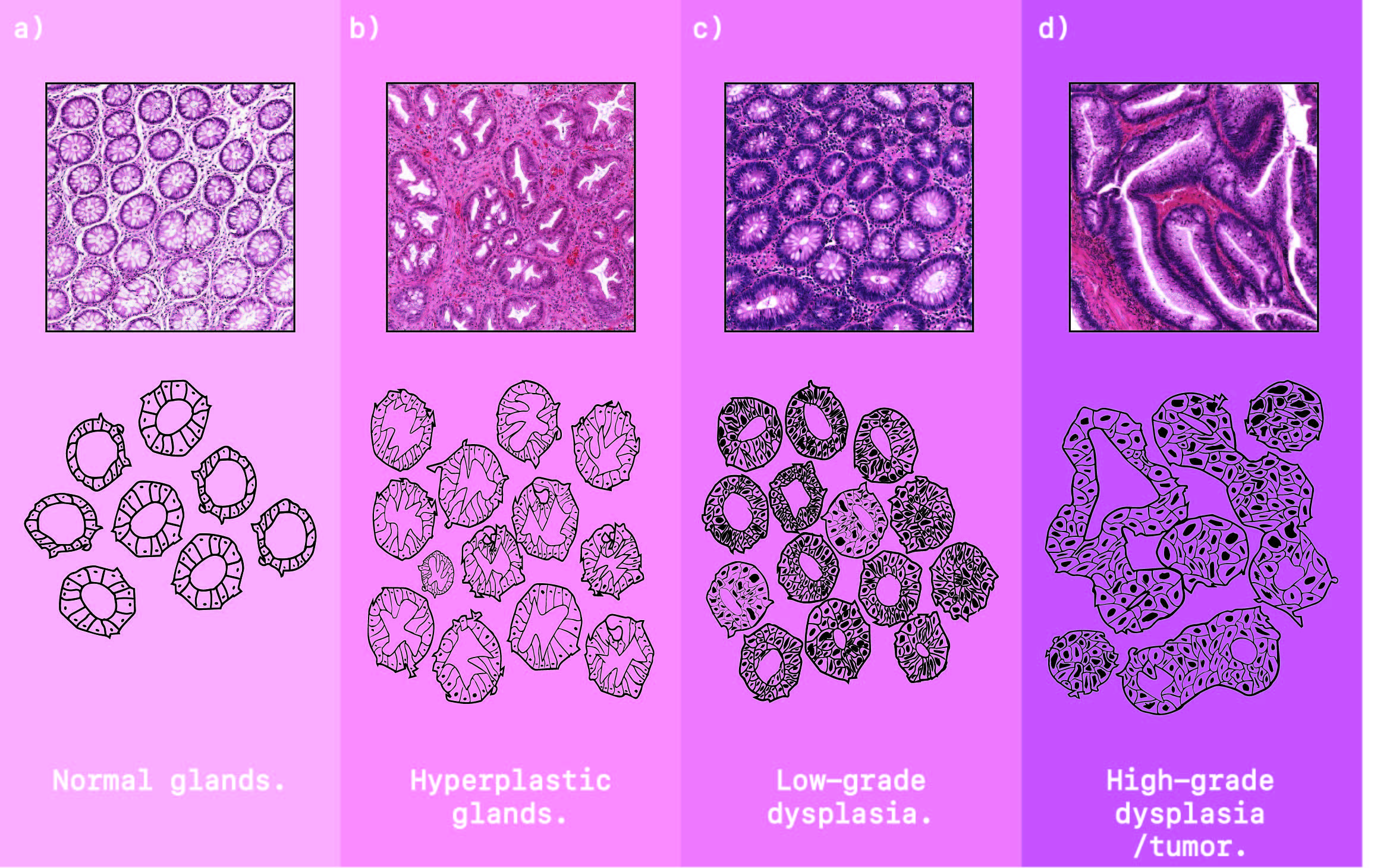}
  \caption{Schematic overview of glandular (de)formation with associated grading class. a) Normal glands; small, organized nuclei and round lumen. b) Hyperplastic gland; small nuclei, saw-tooth like formed lumen. c) Low-grade dysplasia; characterized by unorganized, stacked epithelium cells possibly with enlarged nuclei. d) High-grade dysplasia / tumor; Unorganised fusing glands that oppress the lumen.}
  \label{fig:grades}
\end{figure}

% figure: overall approach
\begin{figure*}[h]
  \centering
  \includegraphics[width=0.8\linewidth]{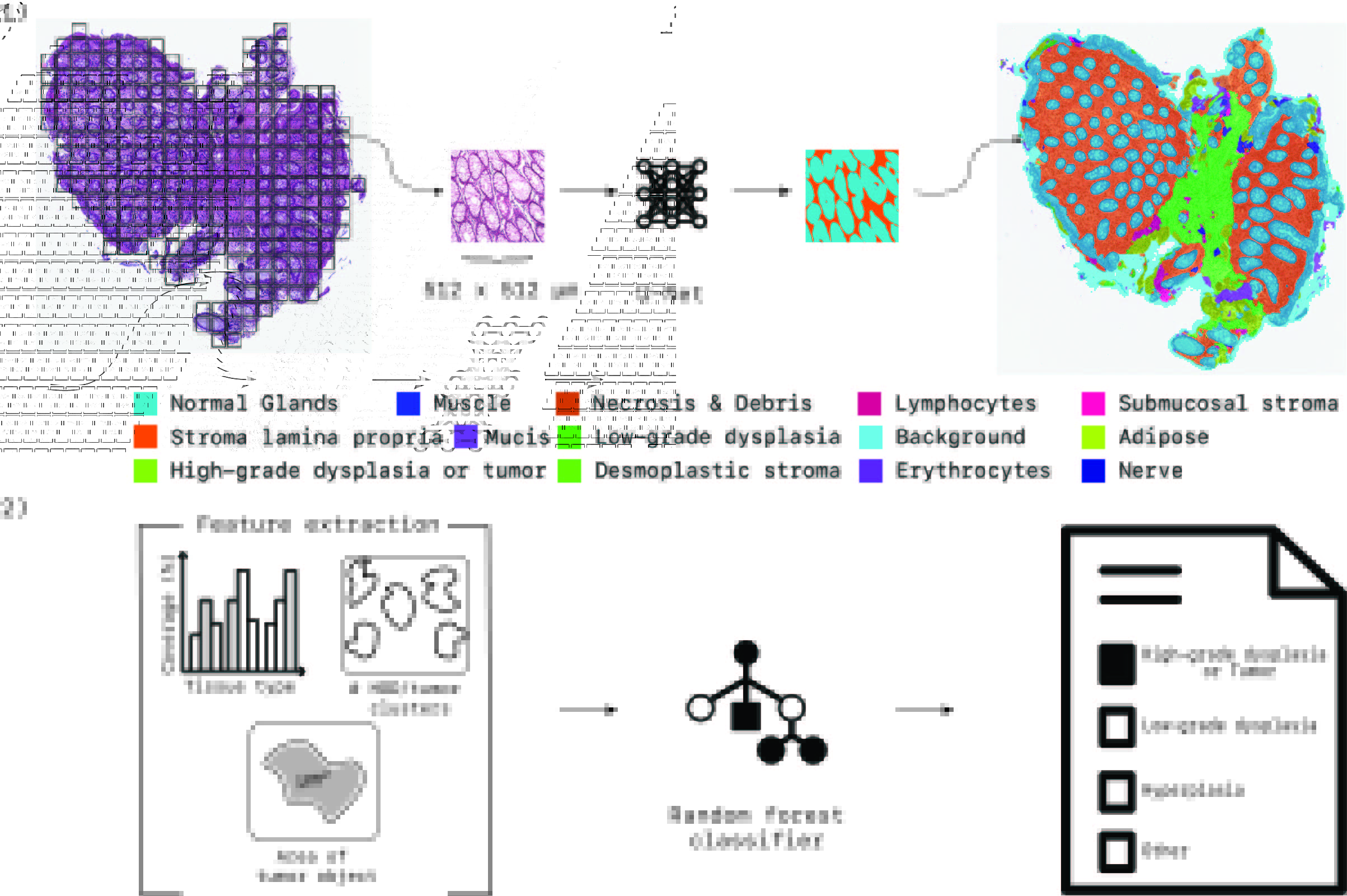}
  \caption{1) Segmentation process; An encoder-decoder segmentation model based on convolutional neural networks processes colorectal tissue tiles with a size of 512x512$\mu$m and segments up to 14 different tissue types. 2) From the segmentation map at slide level, we extract a set of features: a) the normalized histogram of all tissue types, b) the number of high-grade dysplasia / tumor clusters, c) the average, minimum, and maximum size of these clusters. These features are processed by a random forest classifier that gives the final classification.}
  \label{fig:overview}
\end{figure*}

% segmentation
\subsection{Semantic segmentation in histopathology}

% AI in pathology and CRC & related work on AI for segmentation in pathology
In the analysis of medical images such as histopathology whole-slide images (WSI) stained with hematoxylin \& eosin (H\&E), automated analysis of tissue morphology often relies on \emph{segmentation} methods.
With segmentation, an image is divided into a set of non-overlapping regions, each with its particular shape, border and semantic meaning.
Successively, \emph{classification} models can be trained using features extracted from the segmentation output, to make predictions at whole-slide image level \citep{qin18} \citep{naik08}. 
In recent years, deep learning methods based on convolutional neural networks (CNN) have been successfully applied to multiple tasks in the medical domain, showing human-level performance in the field of dermatology \citep{Este17}, radiology \citep{Ardi19}, ophthalmology \citep{Fauw18}, and pathology \citep{bult20}.

In order to diagnose CRC, pathologists differentiate normal epithelial cells from epithelial cells that display characteristics associated with the progression towards cancer.
This process, called \emph{grading}, considers the degree of glandular formation, ranging from normal glands to cancer, with intermediate grades like hyperplasia and dysplasia (see Figure \ref{fig:grades}).
For this reason, a large corpus of deep learning methods has been developed to segment and classify glands (often including their lumen) \citep{bind19}, as well as segmenting all their instances, also fostered by the introduction of international challenges such as GLAS \citep{siri17} and CRAG \citep{grah19}.
%with the top performing networks reaching F1-scores and 0.91 \cite{chen2016dcan} of 0.82 \cite{graham2019mild} on the GLAS and CRAG challenges respectively.
One recent study \citep{wulc21} used tumor segmentation to predict disease free survival for stage II and III colorectal cancer patients.

% state the question about the loss function for segmentation
\subsection{Loss functions for semantic segmentation}
Within the medical image analysis as well as the computational pathology community, CNN models based on an encoder-decoder architecture such as U-Net \citep{ronn15} are nowadays considered the premier choice for image segmentation \citep{bult18} \citep{altu20}.
During training of CNN for segmentation, the parameters of the model are optimized by minimizing the difference, encoded by the \emph{loss function}, between the model's predictions and the real label from the reference standard.
Deciding on which loss function is optimal for a specific segmentation models is still an open question, which is currently being addressed within the computer vision community.
Traditionally, the pixel-wise categorical cross-entropy (CC) has been used in most segmentation models.
However, this loss function performs suboptimally in the presence of over- or under-represented classes in the image, causing the over-represented class to dominate the training of the CNN, leading to a biased model.
In digital pathology, this problem is especially apparent for segmentation tasks where small tissue compartments (e.g., erythrocytes) need to be segmented correctly next to larger components (e.g., muscle). 
To address this, some authors have proposed the use of weights or penalty terms to the CC loss function \citep{long15}.
\citet{xu20b} proposed the use of a \emph{focal loss}, which adds a term to reduce the contribution of easy examples to improve CNN focus on the difficult examples while others \citep{amid19} have made modifications to counteract undesirable effects of noise in the reference standard. 
In conjunction with an increasing use of the \emph{Dice score} and the \emph{Jaccard index} as a metric to assess model performance, differentiable approximations of these two metrics have recently been formulated, known as surrogates, such as \emph{soft-Dice} \citep{sudr17}, \emph{soft-Jaccard} \citep{rahm16} and \emph{Lovasz-softmax} \citep{berm18a}; their differential property makes them usable as a loss function for model training.

% our contributions
\subsection{Our contribution}

% overall
In this paper, we propose to address the growing number of colon biopsies to be analyzed from  population screening by developing a system for automatic classification of colon biopsy WSI based on semantic segmentation of multiple tissue components.
A schematic overview of the proposed solution is depicted in Figure \ref{fig:overview}.

% segmentation + losses
We defined a multi-class segmentation problem including the most representative tissue components in colorectal cancer slides stained with H\&E.
For this purpose, we selected n=14 components, including normal versus low-grade dysplastic (LGD) and high-grade dysplastic (HGD)/cancerous epithelium, stroma lamina propria, submucosal stroma, desmoplastic stroma, muscle, nerve, adipose, mucus, necrosis \& debris and background.
We selected a single encoder-decoder CNN architecture, namely U-Net, and used it to investigate the effect of different loss functions.
In particular, we selected four representative loss functions and compared them: (1) Categorical Cross-entropy loss, (2) Focal loss, (3) Bi-tempered loss and (4) Lovasz-softmax loss.
% development and validation of segmentation + losses
All models were trained using a single-center dataset of n=52 CRC surgical resections or CRC biopsies.
Validation and performance comparison of the different segmentation models using different losses was done on a multi-centric dataset containing n=27 whole-slide images from n=5 different medical centers, as well as on publicly available challenges on CRC segmentation, namely GLAS and CRAG.

% classification
Next, we took the best performing model to a classification stage where we extracted from the segmentation map a number of features based on morphology and used them to train a random forest classifier to assess the risk of biopsies based on four categories used in the pathology report, namely 1) normal, 2) hyperplasia, 3) low-grade dysplasia, and 4) high-risk (high grade-dysplasia or tumor).
% development and validation of classification
We trained and validated our classification system using an external dataset of $>$1,000 patients.

% METHOD
\section{Methods} \label{method}

% Materials
\subsection{Materials}
Two sets of whole-slide images were collected in this study, for two different purposes.
First, a set of cases from colorectal cancer surgical resections and biopsies were collected, with the aim of training and validating multi-class \emph{tissue segmentation} models based on deep learning.
Second, a large set of colon biopsy cases from diagnostic workup were collected, with the aim of developing and validating a multi-class \emph{biopsy classification} model.
In the next sections, both datasets are described in detail.

% segmentation 
\paragraph{Tissue segmentation data}
A total of $n$=79 formalin-fixed paraffin-embedded tissue resection or biopsy specimens of colorectal cancer patients were collected in a multi-centric fashion from four different medical centers in the Netherlands and one medical center in Germany.
All slides were stained with H\&E in the pathology laboratory of each medical center, resulting in a large variety of staining.
Furthermore, all slides were digitized using three different types of digital pathology scanners, resulting in a substantial tissue appearance variation in whole-slide images.
The Pannoramic P250 Flash II scanner (3D-Histech, Hungary) scanner was used to digitize $n$=62 slides provided by Radboud University Medical Center (Nijmegen, Netherlands), $n$=4 slides provided by Eindhoven Medical center (Eindhoven, Netherland), and $n$=5 slides provided by Utrecht University Medical Center (Utrecht, Netherland); 
an IntelliSite Digital pathology slide scanner (Philips, the Netherlands) was used to digitize $n$=5 slides provided by the Leiden University Medical Center (Leiden, Netherland), and a NanoZoomer 2.0 HT (Hamamatsu, Japan) was used for digitizing $n$=3 slides provided by the University of Bayreuth (Germany).
All slides were scanned at a spatial resolution of 0.24 $\mu$m/px and the obtained WSIs were used for model development and validation.

In each WSI, regions of interest covering a broad range of different tissue morphology and tissue components were manually selected.
In each region of interest, one pathologist and two trained medical/technical analysts manually annotated and checked pixels as belonging to one of the 14 following categories: 1) normal glands, 2) low-grade dysplasia, 3) high-grade dysplasia/tumor, 4) submucosal stroma, 5) desmoplastic stroma, 6) stroma lamina propria, 7) mucus, 8) necrosis and debris, 9) lymphocytes, 10) erythrocytes, 11) adipose tissue, 12) muscle, 13) nerve, 14) background.
Each region was exhaustively annotated, meaning that all pixels within the region of interest were labeled, accurately delineating interfaces of different tissue compartments.
Annotations were made using the in-house developed open-source software ASAP\footnote{https://github.com/computationalpathologygroup/ASAP}.
An overview of the proportion of annotated classes compared to the total amount of annotations can be found in Figure \ref{fig:internal_per_results}.
Visual examples of manually annotated regions are depicted in Figure \ref{fig:internal_results}.

A set of n=52 WSIs from a single center (Radboud University Medical Center) and their annotations were randomly selected and used to define a single-center training set (n=40) to develop segmentation models and a validation set (n=12) used to optimize hyperparameters during training.
The rest of WSIs (n=27) and their manual annotations  was used as a multi-centric test-set.
Note that among the n=27 test slides, n=10 were originated in the same medical center as the training set, and the rest (n=17) were originated in different medical centers, and partly scanned using different scanners.

\begin{figure}[t]
  \centering
  \includegraphics[width=0.95\linewidth]{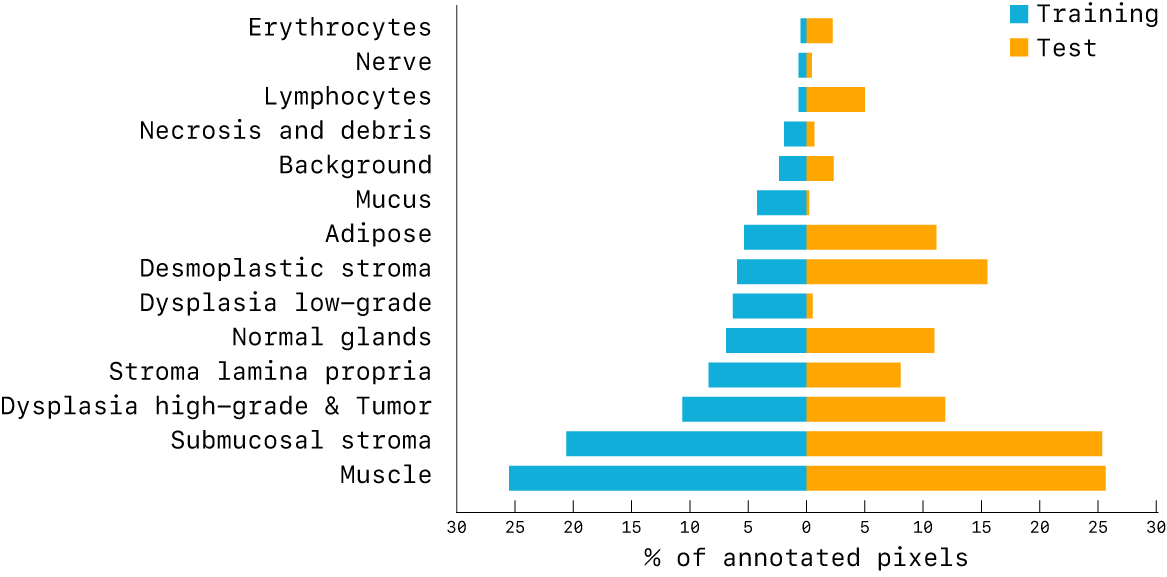}
  \caption{Distribution of the annotated pixels of the training and test-set.}
  \label{fig:annotated_pixels}
\end{figure}

% polyp classification
\paragraph{Biopsy classification data}
A dataset of colon biopsies resected from n=1054 patients was collected from the the digital archives of the Cannizzaro hospital (Catania, Italy).
For each case, the digital pathology slides as well as the related pathology report was collected.
Based on the conclusion of the synoptic report, each tissue sample was assigned by an export to a single label, corresponding to the most clinically relevant finding diagnosed by the pathologist.
We considered most labels that have to be filled-in in the pathology report, which we list here in a decreasing order of clinical relevance: 1) high-grade dysplasia or tumor (n=292), 2) low-grade dysplasia (n=693), 3) hyperplastic (n=36), 4) other (n=33) containing all other classes such as but not limited to normal cases.
When multiple findings were present in a single slide, e.g., both hyperplasia and high-grade dysplasia, the label with the highest associated risk was appointed to the case.
This set of cases was used to develop and validate a prediction model that processes features derived from the segmentation map of each biopsy, and automatically classifies each biopsy into one of the aforementioned categories.
All slides were scanned with a Aperio AT2 (Leica Biosystems) at a spatial resolution of 0.24 $\mu$m/px.

\begin{figure}[t]
  \centering
  \includegraphics[width=1\linewidth]{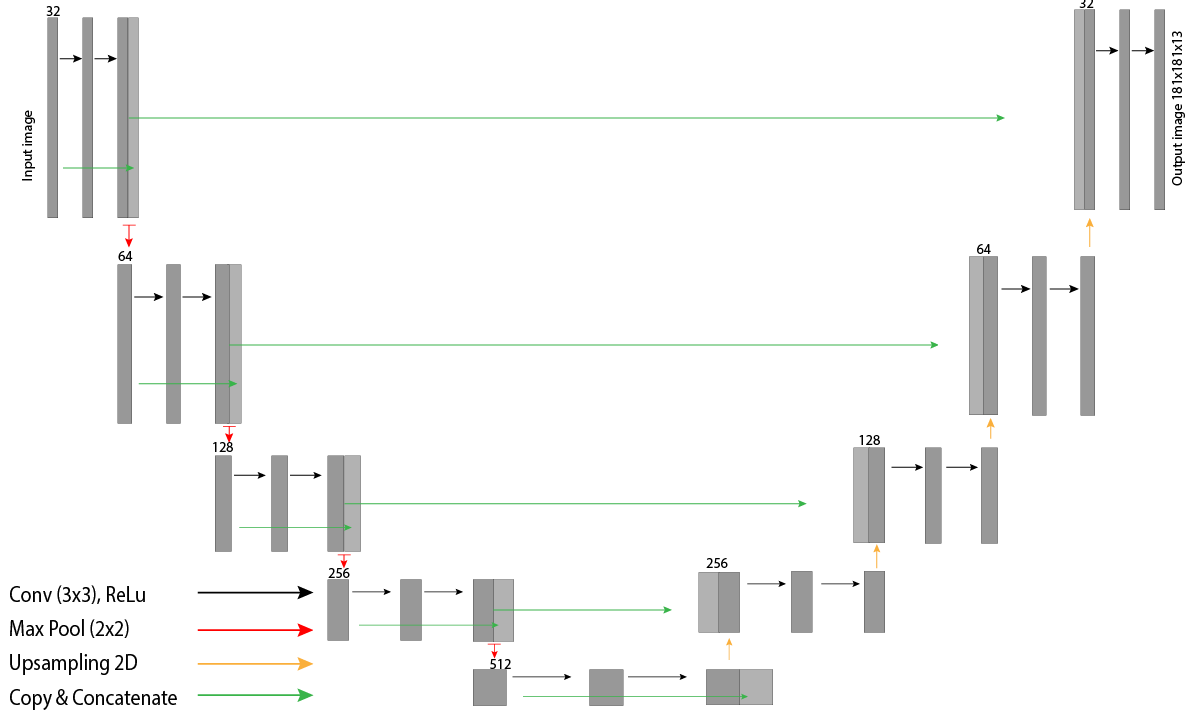}
  \caption{a) Left; example of a WSI with annotation of the training set, Right; example of an WSI with annotations of the test-set. b) Overview figure of the used U-Net architecture. The light-gray boxes represent copies of the feature maps. Each dark-gray box  represents a multi-channel feature map. The input size is shown on the left of the box, and the number of channels on top.The arrows denote the different operations.}
  \label{fig:model_overview}
\end{figure}

\paragraph{Ethical approval}
The use of anonymized data for training and validating the segmentation network has been approved by the Ethical Committee of the Radboud University Medical Center (2015-1637), for training and validation of the biopsy classifier was approved by the Ethical Committee  of the Cannizzaro Hospital (approval number 4428, 12/12/2018). 

% CNN development and validation
\subsection{CNN development and validation}
% In this section, we provide details on the methods proposed in this paper.
First, we detail the CNN model used for automated multi-class tissue segmentation based on the U-Net architecture, followed by the training procedure. 
Secondly, we detail the different loss functions considered and compared in this work, and we provide a short description of each loss function's main characteristics.
Finally, we describe the prediction model developed for the automated classification of biopsies in the context of colorectal cancer screening.

\subsubsection{U-Net for multi-class tissue segmentation}
As done in the original U-Net model, we have doubled the number of filters after every max-pooling layer, but we started with a lower number of filters in the first layer (n=32).
Additionally, inspired by the ResNet \citep{he15b} approach, we introduced additional skip connections within every set of convolutional layers (which we call a U-Net block) before pooling, where the input of the U-Net block is concatenated with the last feature map produced by the block itself.
We observed experimentally that adding these skip connections allowed better flow of the gradients. 
Finally, we replaced transposed convolutions with nearest neighbor up-sampling operations followed by a 2x2 convolution layer in the expansion path.
An overview of the used U-Net model is depicted in Figure \ref{fig:model_overview}.

\subsubsection{Training U-Net for multi-class tissue segmentation}
For training the same settings have been applied to all models.
The input was a RGB patch of 512$\times$512px with a pixel size of 1$\mu$m.
Data were augmented by random flipping, rotation, elastic deformation, blurring, brightness (random gamma), color, and contrast changes during training.
An adaptive learning rate scheme was used, where the learning rate was initially set to 1e-4 and then multiplied by a factor of 0.5 after every 20 epoch if no increase in performance was observed on the validation set.
The weights of the network were initialized as proposed in (He et al., 2015).
The mini-batch size was set to 5 instances per batch, the networks were trained for a maximum of 500 epochs, with 300 iterations per epoch. 
Training of the networks was stopped when no improvement of the validation loss was found for 50 epochs. 
The output of all networks is in the form of $C$ likelihood maps.
The arg-max was taken to obtain a final segmentation output.

\subsubsection{Loss functions for multi-class tissue segmentation}
\paragraph{Categorical cross-entropy}
In the context of image segmentation, the cross-entropy loss, also known as softmax loss, has been largely used both in binary and multi-class problems, referred to as categorical cross-entropy (CC) loss in the presence of a multi-class problem.
This loss compares how well the probability distribution output $\hat{y}$ of the final softmax layer of a deep neural network matches the value of the ground-truth data $y$.
In segmentation tasks, this is done pixel-wise for every pixel in the image.
Typically, the sum or mean of all values together gives the final single value for the entire input image. 
The CC loss is computed as
\begin{equation}
% \begin{align}
    loss = - \sum_{i}^{C} y_i \log \hat{y}_i,
% \end{align}
\end{equation}
where $y_i$ and $\hat{y}_i$ are the ground truth and the softmax output of the model for each class $i$ in $C$. 

Although being one of the standard loss functions, it is known to perform poorly under certain conditions (i.e. class imbalance or noise in the dataset). 

\paragraph{Focal loss}
Many authors have been concerned with the problem of imbalanced classes: negative examples significantly outnumber positive examples, and the huge number of background examples (or easy-negative examples) overwhelms the training.
Training a network with the CC loss on an imbalanced dataset can lead to a network that is biased towards the data-dominated classes. 
This becomes especially apparent in segmentation tasks when small objects need to be segmented. 
Because of the low pixel count, these small objects don't have a large contribution to the loss function.  
To overcome this issue, \citet{Lin17a} proposed the Focal loss:
\begin{equation}
% \begin{align}
    loss = - \sum_{i}^{C} \alpha (1 - \hat{y}_i)^\gamma y_i \log \hat{y}_i
% \end{align}
\end{equation} 

Presented initially to improve single-shot detection networks (such as YOLO \citep{redm17}, SSD \citep{liu16}, and Retinanet \citep{Lin17a}), it has shown to work for segmentation problems as well.
This loss modifies the cross-entropy loss, so objects/pixels that are 'easy' to classify or are present abundantly are weighted lower, resulting in a smaller loss value.
More complicated objects/pixels are weighted heavier, resulting in a higher loss value.
How much up- or down-weighted the values are, is determined with two additional hyperparameters ($\alpha$ and $\gamma$).
The authors have tested different values of the two hyperparameters in the original paper and suggested two optimal values.
Given the scope of this paper, we have simply adopted both values here.

\paragraph{Bi-tempered loss}
The presence of small errors, flaws or defects (noise) in the dataset, for example as a result of small inaccuracies in the annotation can hamper the quality of a segmentation output disproportionately.
In such scenarios, the CC loss value can grow without boundary as these outliers can be far away from any decision boundary: the model is penalized by noise in the dataset.
To compensate for this, the network stretches its decision boundary, resulting in a less robust model.
To counteract this effect, \citet{amid19} introduced the Bi-tempered loss function. 
They propose two changes to the default softmax CC. 
First they replace the softmax output with a heavy tailed softmax, that acts as a form of 'label smoothing'. 
The replacement softmax function is given by: 
\begin{equation}
% \begin{align}
\begin{split}
    \hat{y}_i = \exp_{t_2} (\hat{a}_i - \lambda_{t_2}(\hat{a})), where \ \lambda_{t_2}(\hat{a}) \in \mathbb{R} 
\end{split}
% \end{align}
\end{equation} 
such that $\sum_{j}^{C} \exp_{t_2} \left( \hat{a}_j  - \lambda_{t_2}(\hat{a}) \right) = 1$,

where $a$ are the activations from the final layer of the network, and $a_i$ is the output for class $i$, $t_2$ is a hyperparameter. 
The second change they make is by replacing the entropy function with a tempered version, given by: 
\begin{equation}
% \begin{align}
    loss = \sum_{i}^{C} \left( y_i(\log_{t_1} y_i - \log_{t_1} \hat{y}_i) - \dfrac{1}{2-{t_1}}(y_i^{2-{t_1}} - \hat{y}_i^{2-t_1}) \right).
% \end{align}
\end{equation} 
The two hyperparameters determine how heavy tailed the functions become. 
When both hyperparameters ${t_1}$ and ${t_2}$ are set to 1, the standard logistic loss is recovered. 
We have not tested this option either, but have kept the default values proposed by the authors.

\paragraph{Lovasz loss}
Optimization for cross-entropy entails a problematic relationship between the learning optimization objective (the loss) and the end target metric.
Therefore, recent works in computer vision have proposed soft surrogates to alleviate this discrepancy and directly optimize the desired metric, either through relaxations (soft-Dice, soft-Jaccard/Intersection over Union) or submodular optimization (Lovasz-softmax).
Dice loss is based on the Sorensen-Dice coefficient, which attaches similar importance to false positives and false negatives, and is immune to the data-imbalance problem. 
Instead of a pixel-wise approach, the Dice or Intersection over Union (IoU) calculates the similarity between two samples of the same class resulting in a value between 0 and 1.
The scores are often averaged over all classes for multi-class problems, resulting in the mean-Dice or mean-IoU.
Advantages of this index compared to per-pixel losses are scale invariance and appropriate counting of false negatives, although it tends to favor large objects over small objects.
To deal with this issue \citet{Berm18b} combined the Lovasz hinge with the IoU resulting in a loss function that seems to focus both on small and large objects.

\begin{equation}
    loss = \dfrac{1}{\mid C \mid}  \sum_{i}^{C} \overline{\Delta_{J_c}} \left( \textbf{m}(i) \right)
\end{equation} 

Where $ \overline{\Delta_{J_c}} $ is the Lovasz hinge of the IoU and $\textbf{m}(c)$ the class probabilities for class $c$. 

\paragraph{Dice loss}
% class balancing from Bokhorst et al., 2019
In a previous study \citep{Bokh19}, we have studied and tested a number of different class balancing methods.
The Dice loss proved difficult to implement in combination with our preferred class balancing method.
We therefore chose to maintain / re-apply the relevant class balance method in this study and not to include the Dice loss in the comparative study.
We will use the Dice metric however, to evaluate the models performance.

\subsection{Biopsy classification}
An additional classification step is needed to obtain a single label per slide from a segmentation output.
We have opted for a simple random forest classifier for this, combined with five easy interpretable features.
We extracted the number of pixels of every class in the segmentation output (histogram) and normalized it by dividing the histogram by the total number of segmented pixels as the first feature. 
The normalized histogram does not tell us anything about the spatial relationship between these pixels though; for example, they could be scattered all over the tissue as noise. 
Since tumor mostly comes in clusters we, in addition to the normalized histograms (1), used connected components to obtain the number of tumor clusters per case (2), along with the average (3), min (4), and max (5) size of all clusters.
We experienced that the segmentation output sometimes incorrectly segmented small clusters as tumor. 
Therefore, we excluded all clusters smaller than 30 $\mu$m$^2$ from the segmentation output to correct these mistakes.
The exclusion cut-off value of 30 $\mu$m$^2$ was found empirically.

A random forest with 1000 decision trees was trained to produce the final single label output.
The random forest was trained in a five-fold cross-validation setup.
Before training, all feature values were normalized to have zero mean and scaling to unit variance.
The same normalization parameters were applied to the test-set before running the classifier.  
Per fold, a random forest was trained using four of the five folds as training data.
The performance of the model is validated on the remaining part that was left out during training. 
We evaluated the performance of the trained model with a 1-vs-all Receiver Operating Characteristic (ROC) analysis. 

\section{Results}

% segmentation results on patches: 1) radboud, 2 bayreuth, 3 leiden, 4 utrecht (project....), 5 eindhoven (Cream)
\begin{figure*}[t]
  \centering
  \includegraphics[width=0.95\linewidth]{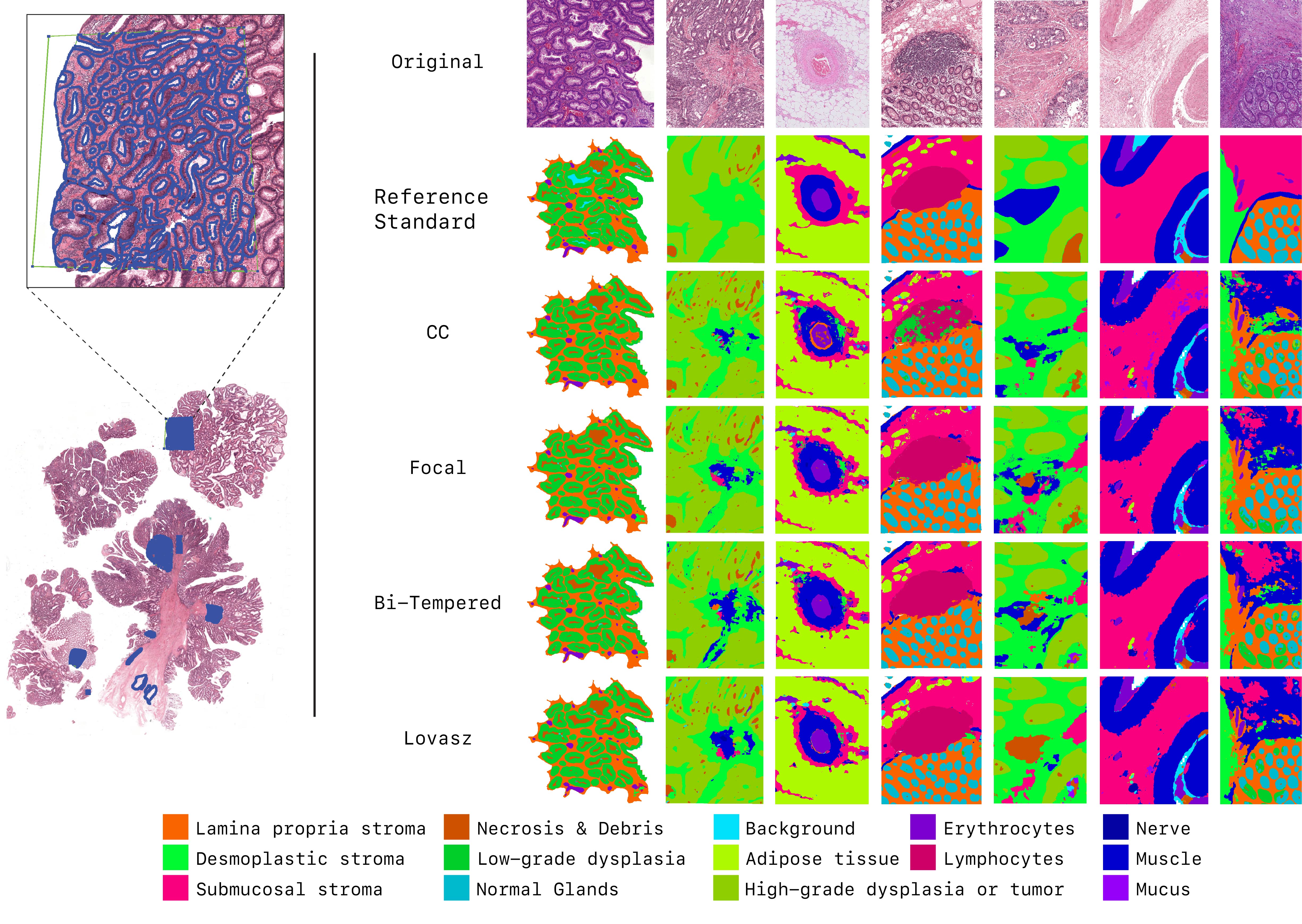}
  \caption{Left; example of manual annotations. Right; segmentation output on the private test-set. Every image shown is from another institute from the test-set.}
  \label{fig:internal_results}
\end{figure*}

% Internal validation
\subsection*{Segmentation performance on a private multi-centric dataset}
% Total number of annotation regions = 165, max size = 4.08mm2, min_size = 0.1, 999 number of non overlapping tiles
We evaluated the different loss functions' effect on 999 non-overlapping tiles of 512x512$\mu$m extracted from a total of 27 WSI from 5 different centers, hereafter referred to as private test-set. 
The non-overlapping tiles were extracted from 165 manually annotated regions with an area between 0.05 mm$^2$ and 4.08 mm$^2$.  
The Dice-score was selected to evaluate the performance of the different models. 
Per model, we calculated an overall and an individual (class) Dice-score (see Table \ref{dice_alltest}). 

The model trained with the Lovasz-softmax loss achieves the overall best performance with a Dice-score of 0.72, although this score is well matched by the Bi-tempered loss model (Dice-score = 0.71). 
Models trained with the CC and Focal loss score slightly lower (Dice-score = 0.69). 
The Wilcoxon signed-rank test shows there is no significant difference between the overall scores of the loss functions. 

Although the differences between the overall scores of models trained with different loss functions are marginal, there are sometimes notable differences for the results per class. 
For example, the Bi-tempered loss outperforms the other loss functions on segmenting low-grade dysplasia (Dice-score = 0.88) and similar forms of specialization can be found for the other classes/loss functions. 
Results are depicted in Figure \ref{fig:internal_results}.

The overall Dice-scores per test-center for the centers in the private dataset are in line with the overall average Dice-scores. 
However, in one of the external test-centers, all trained models show suboptimal performance in the presence of submucosal stroma, presenting a drop from an averaged  Dice-score of 0.54 overall to 0.28. 
Further investigation of this test-center shows that these slides have a very dark H\&E staining that could harm the performance of the networks.
An overview of all Dice-scores per center can be found in Appendix \ref{fig:internal_per_results}. 

\subsection*{Segmentation performance on public datasets}

\begin{figure*}[t!]
  \centering
  \includegraphics[width=0.80\linewidth]{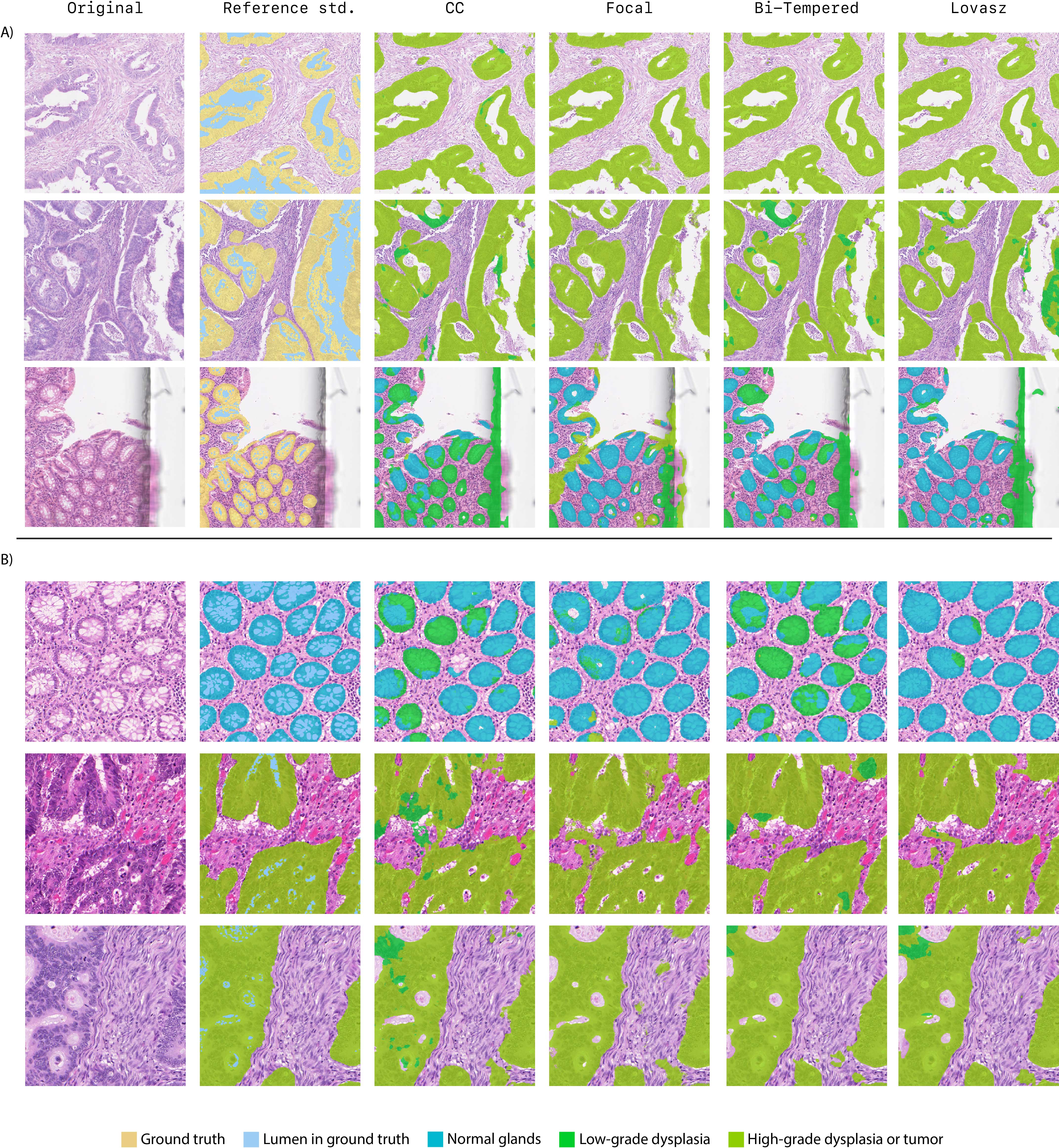}
  \caption{Segmentation output on the A) CRAG and B) GLAS challenge. The F1-scores are calculated on the reference standard images where the background (light-blue) has been removed. Note, because the CRAG challenge is a binary segmentation task, we marked the epithelium in the reference standard with a single color. }
  \label{fig:crag_results}
\end{figure*}

\begin{figure*}[t]
  \centering
  \includegraphics[width=0.90\linewidth]{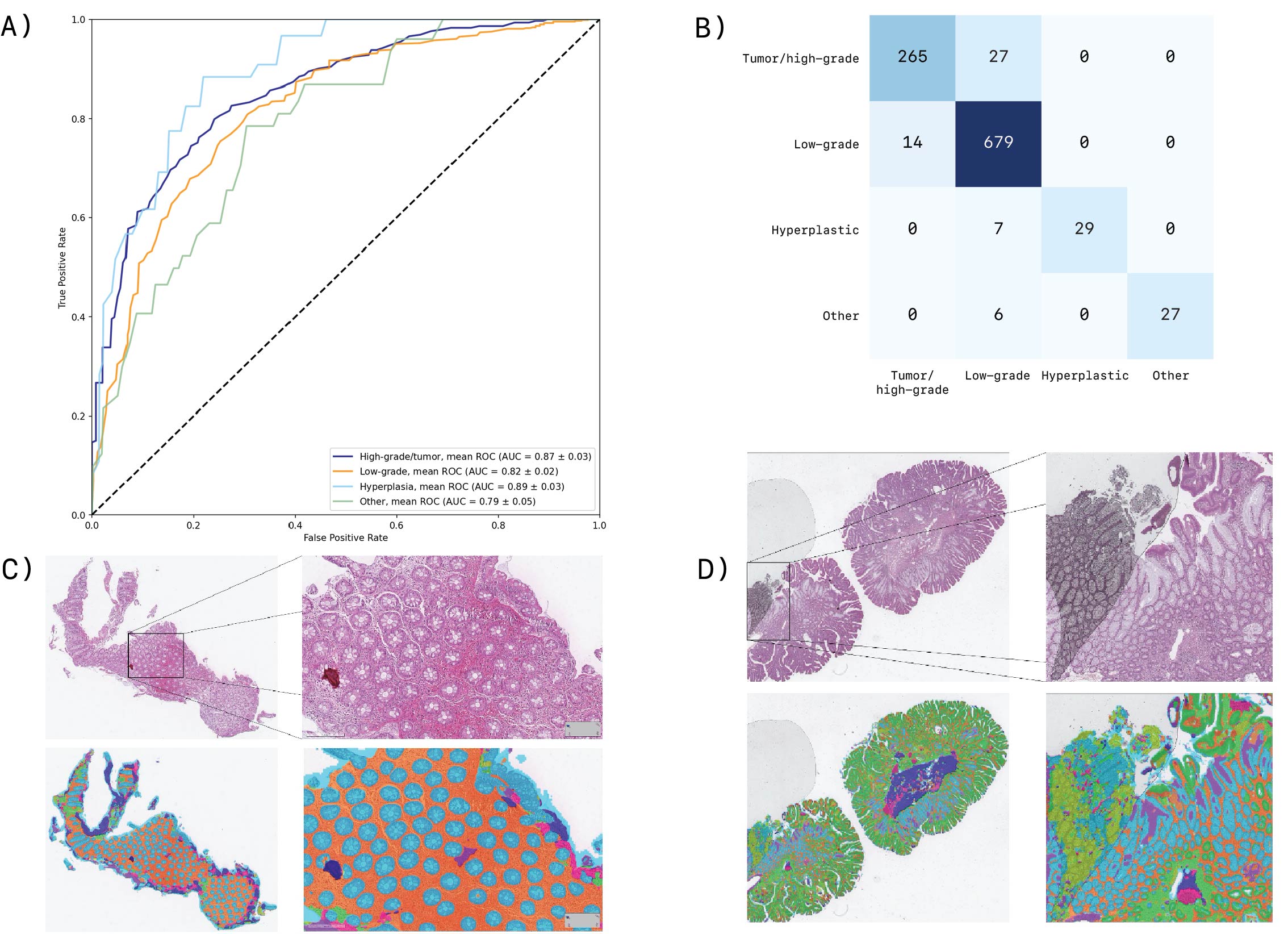}
  \caption{A) ROC-curves of the random forest classifier. B) Confusion matrix of the biopsy classifier. C) Segmentation output of a healthy fragment. D) Segmentation result that fails due to air bubble on the glass slide.}
  \label{fig:biopsy_results}
\end{figure*}

\begin{table}[t]\small
\centering
\begin{tabular}{|c|c|c|c|c|}
\hline  & \textbf{CC} & \textbf{Focal} & \textbf{Bi-tempered} & \textbf{Lovasz}  \\ \hline
\textbf{Tumor}                 & \textbf{0.89} & 0.87  & 0.87  & 0.83  \\ \hline
\textbf{Desmoplastic Stroma}   & \textbf{0.69} & 0.54  & 0.64  & 0.58  \\ \hline
\textbf{Necrosis and Debris}   & 0.46  & \textbf{0.49} & 0.47 & 0.45   \\ \hline
\textbf{Lymphocytes}           & 0.61  & 0.82  & 0.82  & \textbf{0.84} \\ \hline
\textbf{Erythrocytes}          & 0.62  & 0.66  & 0.65  & \textbf{0.68} \\ \hline
\textbf{Muscle}                & 0.81  & 0.84  & \textbf{0.85} & 0.82  \\ \hline
\textbf{Submucosal Stroma}     & 0.50  & \textbf{0.58} & 0.55 & 0.54   \\ \hline
\textbf{Adipose tissue}        & 0.85  & 0.85  & \textbf{0.89} & 0.86  \\  \hline
\textbf{Mucus}                 & \textbf{0.64}  & 0.46  & 0.60  & 0.62  \\ \hline
\textbf{Nerve}                 & 0.63  & 0.54  & 0.69  & \textbf{0.83}  \\ \hline
\textbf{Normal Glands}         & 0.85  & 0.87  & 0.82  & \textbf{0.88}  \\ \hline
\textbf{Stroma lamina propria} & \textbf{0.88} & 0.86  & 0.87 & 0.82    \\ \hline
\textbf{Background}            & 0.46 & 0.45 & 0.36  & \textbf{0.51}     \\ \hline
\textbf{Dysplasia low grade}   & 0.80 & 0.78 & \textbf{0.88}  & 0.77     \\ \hline
                               &      &      &                &          \\ \hline
\textbf{Average}               & 0.69        & 0.69           & 0.71                 & \textbf{0.72}                    \\ \hline
\end{tabular}
\caption{Dice-scores of different loss functions on the entire private test-set.}
\label{dice_alltest}
\end{table}

\begin{table}\small
\begin{tabular}{|c|c|c|c|c|}
\hline
\textbf{} & \multicolumn{2}{c|}{\textbf{CRAG}} & \multicolumn{2}{c|}{\textbf{GLAS}} \\ \hline
\textbf{} & \textbf{w lumen} & \textbf{w/o lumen} & \textbf{w lumen} & \textbf{w/o lumen} \\ \hline
\textbf{CC} & \textbf{0.68} & \textbf{0.77} & \textbf{0.79} & \textbf{0.80} \\ \hline
\textbf{Focal} & 0.66 & \textbf{0.77} & 0.76 & 0.78 \\ \hline
\textbf{Bi-tempered} & 0.54 & 0.69 & 0.71 & 0.75 \\ \hline
\textbf{Lovasz} & \textbf{0.68} & 0.76 & \textbf{0.79} & \textbf{0.80} \\ \hline
\end{tabular}
\caption{F1-scores of different loss functions on the public CRAG and Glas datasets. Results are computed both when the luminal area (w lumen) is included as well as removed (w/o lumen).}
\label{tab:dice_crag}
\end{table}

The CRAG challenge's goal is to segment all glands (epithelium tissue, lumen included), either with or without pathological change. 
As our network also differentiates between different epithelial conditions, we combined our classes 'normal epithelium', 'low-grade dysplasia', and 'high-grade dysplasia/tumor', for the purpose of comparison with this dataset. 
Because the GLAS challenge distinguishes between benign and malignant epithelium, we combined our classes 'normal epithelium' and 'low-grade dysplasia' for comparison in that context. 
For these comparisons, we further relabelled the lumen (pixels with a mean RGB value higher than 240) as background in both public datasets since our networks are designed to segment such regions separately. 
Because our network is more precise in these regions, we report both results with and without lumen, to show the impact of this on our models' performance.

The results on the GLAS and CRAG datasets deviate from the results on the private test-set. 
The CC and Focal losses show increased performance on the CRAG dataset (without lumen) with a Dice-score of 0.77, and Lovasz-softmax loss performs slightly worse here (Dice-score = 0.76), but the aforementioned three loss functions perform almost equally well on the GLAS set (without lumen).
Remarkably, the Bi-tempered model achieves the worst results on the two public data sets with Dice-scores of 0.69 and 0.75 on CRAG and GLAS, respectively. 

Without the removal of the lumen, all networks obtained lower Dice-scores. 
On the CRAG dataset, all networks show Dice-scores that are on average 0.1 lower. 
The CC and Lovasz-softmax loss perform best with a Dice-score of 0.68, followed closely by the Focal loss with a Dice-score of 0.66, the Bi-tempered loss obtained a Dice-score of 0.54. 
On the GLAS dataset, the differences with and without lumen are less evident. 
The CC, Focal, and Lovasz-softmax loss obtain almost equal Dice-scores of 0.79, 0.76, 0.79, respectively. 
The Bi-tempered loss achieves a Dice-score of 0.71. 
The Dice-coefficients are shown in Table \ref{tab:dice_crag}. 
Visual results on Glas and CRAG data can be found in Figure \ref{fig:crag_results}.

\subsection*{Evaluation of the colon biopsy classifier}
Biopsy classification relies not only on the segmentation of the epithelium classes but also on surrounding stromal layers. 
Therefore, the segmentation network should be capable of segmenting all tissue-types correctly. 
Since the network trained with the Lovasz-softmax loss obtained the overall best Dice-scores and seemed to be the most stable across all different classes, this model was selected for the biopsy classification. 
One thousand fifty-four biopsies, including polyps, were collected from the Cannizzaro hospital's pathology department in Catania (Italy) and were processed by the model trained with the Lovasz-softmax loss. 
We collected the normalized histogram, the total number of tumor clusters, average/min/max size of all tumor clusters, and every tissue-fragment within the WSI from the segmentation output. 
Five-fold cross-validation was used to train and validate the random forest classifier. 
We report the aggregated results on the left-out cases.
Before training, all features were normalized to have zero mean and scaled to unit variance. 
The same statistics learned from the training set were applied to the test-set before using the classifier. 
A WSI can contain multiple sections of a biopsy. Consequently, it is not guaranteed that all tissue fragments contain the same morphology. 
Therefore, all WSI tissue fragments are processed; the worst overall grade is then selected as the final slide label.
As shown in figure \ref{fig:biopsy_results} we found an AUC of 0.87 ($\pm 0.03$), 0.82 ($\pm 0.02)$, 0.89 ($\pm 0.03)$, and 0.79 ($\pm 0.05)$ for the classification of HGD/tumor, low-grade dysplasia, hyperplasia, and other respectively. 
Misclassification of healthy tissue by labeling it as tumor tissue is worse than mislabeling healthy tissue as hyperplastic tissue. 
Therefore, we used the quadratic weighted kappa to evaluate the classifier's results. 
An overall kappa score of 0.91 was reached. 
The confusion matrix for all cases and some examples of the segmentation output of the biopsies can be found in Figure \ref{fig:biopsy_results}.

\section{Discussion}
\begin{figure*}[t]
  \centering
  \includegraphics[width=0.95\linewidth]{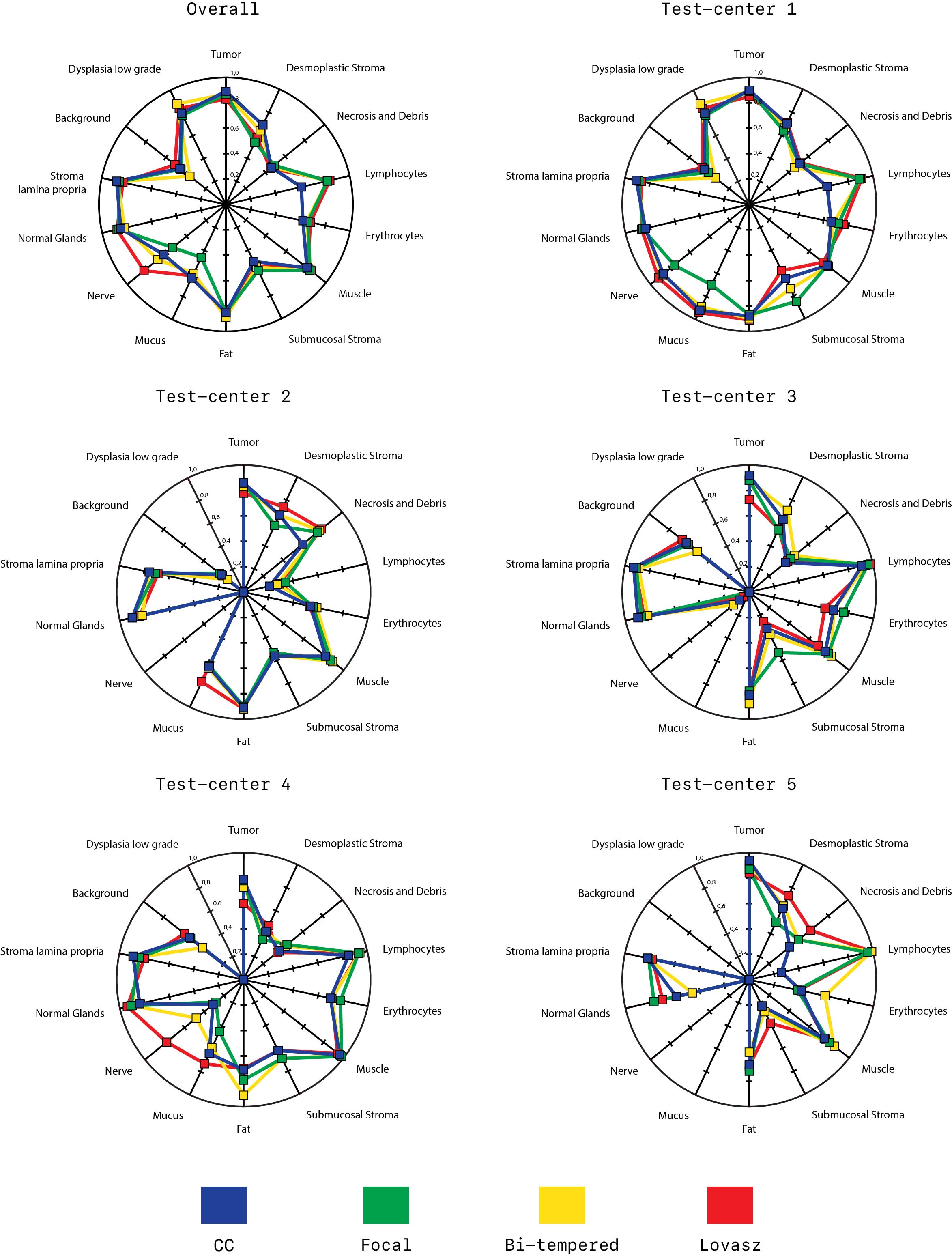}
  \caption{DICE scores of the different loss functions per class of; top-left) the entire test-set or per test-center in the private dataset. Note that if a specific class is not present the DICE-score is zero.}
  \label{fig:internal_per_results}
\end{figure*}

% intro on what we have done: segmentation + classification?
When developing a deep learning model, several components determine the quality of the final result. 
This paper has set out to determine the effects of different loss functions in a 14-class semantic segmentation task. 
We selected four different state-of-the-art loss functions, each with its own characteristics, and trained them on WSI's of a single center. 
The four models' performance on all fourteen classes can best be judged based on the private test-set. 
Although the differences are not significant, the Lovasz-softmax performs best overall on the considered segmentation task. 
The CC-loss performance lags somewhat behind this loss, the same applies to the focal loss, which also shows no added value compared to the CC on the private data set.
However, it is conceivable that the potentially positive effect of the focal loss is lost with a large / larger number of classes. 
The Bi-tempered loss almost equals the performance of the Lovasz-softmax loss on the private test-set and distinguishes itself positively from both the CC and the focal loss on this set. 
We note that this model mainly underperforms on the background class, although all networks have trouble with this class, in addition to the necrosis class. 
We observe some mistakes between the different epithelium classes in some of the individual test-centers of the private test-set, as can bee seen in Figure \ref{fig:internal_per_results}. 
We tend to attribute this effect to stain variations in certain classes which we will discuss further below. 

The Bi-tempered loss is performing worse on the public test-set, despite the relatively high dice scores on epithelium classes (healthy glands, low-grade dysplasia, high-grade dysplasia/tumor) on the private test-set. 
Images from the Bi-tempered results on the public sets show over-segmentation of the high-grade dysplasia/tumor classes, which can explain these results. 
Both the Focal and the Bi-tempered loss have two hyperparameters that can influence the overall quality of the networks. 
For this study, we used the values that the authors of the original papers recommended. 
Further improvements can be made by searching for the optimal parameters but out of this paper's scope.

The Lovasz-softmax loss manages to maintain its good performance on public test-set and seems to benefit from its relative scale insensitivity in this classification task. 
We think that this also results in an overall smoother representation of the output compared to the other loss functions whose images look a bit grainier.
Clearer border definitions were also observed by the WSI's processed with the Lovasz loss compared to the other networks. 
Although there is no significant difference between the metric sensitive Lovasz-softmax loss and the other losses on the various datasets, the lovasz shows a good and stable prediction quality on the private and public datasets as well.
With the additional benefit of no additional hyperparameter tuning. 
This result is in line with \citet{eelb20}, who not only demonstrate the mathematical superiority of multiple metric-sensitive solutions compared to the CC (inspired) loss functions, but also substantiate this with empirical research.

When designing the classification model, we have chosen to stay as close as possible to the result of the segmentation network, assuming that the output of an accurate segmentation model, next to being fully interpretable by pathologists, needs little engineering to generate discriminative features to stratify the risk of colon biopsies.

For this reason, the relative amount of tissue per class (class histogram) has been introduced as a first feature.
Because this factor lacks spatial distribution information, we decided to include such information for the epithelium class 'tumor' in the model, in order to be able to more accurately separate this class from related (epithelial -) classes. 
To this end, we selected the number of tumor clusters, the minimum, maximum and average cluster size as an additional feature.
We have chosen to apply our segmentation model with Lovasz-softmax optimization function to 1054 biopsies. 
From the segmentation outputs we extracted the features and trained a simple random forrest. 
We achieve a quadratic kappa score of 0.91 on this independent dataset. 

We performed an extensive error analysis with the involvement of pathologists.
This analysis showed that roughly 40 \% of the classification errors can be traced back to a faulty output of the segmentation model. 
In overstained specimens, the dark regions parts are sometimes incorrectly classified by our segmentation model as tumors. 
This effect can be addressed by a more substantial stain augmentation and a more extensive targeted training on dark-stained data but requires retraining the network.
An alternative would be to apply stain normalization during inference for which no retraining is required. 
An example of this is the use of Cycle-GANS as propised by \citet{bel21}. 
Other sources of errors are related to incidental artifacts such as small air bubbles in the WSI’s and partly with staining (artifacts) (as shown in Figure \ref{fig:biopsy_results}d).
Suboptimal performance due to these kind of artifacts can be countered by adopting some form of quality control of digital slides, or by explicitly introducing regions with artifacts in the training set.
The other errors cannot be traced back to faulty segmentation but are related to the classifier's limitations. 
An example of this are more difficult cases when for example low-grade and high-grade dysplasia/tumor are present in the same biopsy. 
Since information about low-grade dysplasia is only present in the normalized histogram, we expect that adding more features about these classes could improve the classification output. 

The proposed model has shown good performance of biopsy classification of whole-slide images of colorectal tissue. 
Because the classification model uses easy to interpret features based on a segmentation model, it is easy for pathologists to understand the results of the models.
Future work would include validation on a larger cohort from different centers around the world, in addition to investigating additional features.

\section{Conclusion}
In this paper, we have compared performance of semantic segmentation models for histopathology images in CRC using four different loss functions, three of which are per-pixel CC (related) and one metric sensitive loss (Lovasz-softmax loss). 
All networks were trained on a single center dataset and evaluated on three medical segmentation tasks from multiple centers. 
We found no major differences between the performance of the different loss models, but saw a consistently best performance of the Lovasz-softmax function on all tasks and a variable task-dependent prediction quality of the Bi-tempered loss. 
We definitely see the use of the Lovasz-softmax loss as the better alternative for both the CC and the Focal loss, but also for the Bi-tempered loss, provided there is no significant degree of noise in the dataset.
In practice the Lovasz-softmax performs equally or better than the other losses, and visually gives a more accurate, and cleaner segmentation result, and a better definitions of the borders. 
Since there are no hyperparameters to tune, it is easier to use in comparison to the Focal and Bi-tempered loss. 

We used the model trained with the Lovasz-softmax loss to segment a series of more than thousand biopsies and classified them into four classes, in line with current pathology reports, using a simple classifier and a handful of features, derived directly from the segmentation output. 
We showed that with a good segmentation as base one can obtain very good results on a downstream task. 
The classifier could potentially support pathologists in diagnosing colon biopsies in the context of population screening. 

\section{Code availability}
The segmentation model is available for research use on \url{https://grand-challenge.org/algorithms/colon-tissue-segmentation/}. 

\section{Acknowledgements}
This project has received funding from the Dutch Cancer Society, project number 10602/2016-2, from the Alpe dHuZes / Dutch Cancer Society Fund, grant number KUN 2014-7032, and from the European Union's Horizon 2020 research and innovation programme under grant agreement No 825292 (ExaMode, htttp://www.examode.eu/).

\section{Contributions}
J-M.B. Conceptualization, Methodology, Software, Validation, Data Curation, Writing - original draft,  Visualization. F.P., S.M., W.M. and M.V.  Resources and Writing - Review \& Editing . I.N., J.vd.L. and F.C. Conceptualization, Supervision, Writing - Original Draft , and Writing - Review \& Editing.

%%Harvard
\bibliographystyle{model2-names.bst}\biboptions{authoryear}
\bibliography{bokhorst19}

\section*{Supplementary Material}

\subsection{Overview of DICE-scores per center}
\end{document}